\documentclass[aps,prb,twocolumn,groupedaddress]{revtex4}
\usepackage{graphicx}
\begin{document}


\title{A spaceship with a thruster - one body, one force}


\author{Scott C. Dudley and Mario A. Serna}
\affiliation{U.~S.~Air Force Academy, Colorado Springs, CO  80840}


\date{\today}

\begin{abstract}
A spaceship with one thruster producing a constant magnitude force
is analyzed for various initial conditions.  This elementary
problem, with one object acted upon by one force, has value as a
challenge to one's physical intuition and in demonstrating the
benefits and limitations of dimensional analysis.  In addition,
the problem can serve to introduce a student to special functions,
provide a mechanical model for Fresnel integrals and the
associated Cornu spiral, or be used as an example in a numerical
methods course.  The problem has some interesting and perhaps
unexpected features.

\end{abstract}

\pacs{}

\maketitle

\section{Introduction}
A problem involving one constant magnitude force acting on one
body leads to interesting motion and is useful as a teaching tool
when discussing physical intuition and dimensional analysis.  The
problem can be solved with elementary mechanics with the solution
expressed in the form of integrals associated with known special
functions.  We first state the problem and then proceed with its
solution, though we invite the reader to ponder the problem before
reading the solution and the remainder of the article.  After the
solution is presented we discuss what one could have ascertained
from an astute application of physical intuition.  Next,
dimensional analysis is applied to the problem. We hope
instructors will find the richness of such a simple to state
problem to be of value in teaching and in challenging students. In
addition, the problem can be attacked analytically to a large
degree and thus comparison of analytic (and asymptotic
expressions) to numerical solutions could be instructive in a
numerical methods course.

\section{The Problem}
\begin{figure}
\includegraphics[width=3 in]{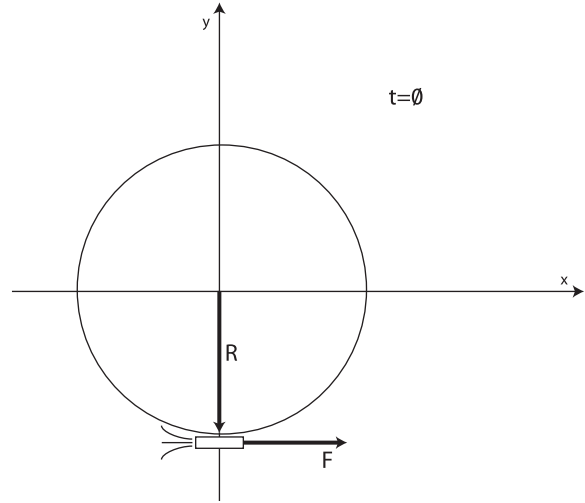}
\caption{\label{fig:spaceship} Position of spaceship/thruster
combination at time $t=0$.}
\end{figure}

Imagine a spaceship with one thruster positioned a distance $R$
from the center of mass as depicted in Fig.~\ref{fig:spaceship}.
Initially the ship is at rest. At time $t=0$, the thruster is
fired and produces a constant tangential force, $F$. Describe the
motion of the ship. What path does the center of mass move along?
Assume special relativity is not needed and that the mass of the
spaceship/thruster combination does not change. What else can one
intuit about the motion?  The solution follows immediately so we
suggest the reader formulate opinions about the motion before
proceeding.


\subsection{The solution to the problem} 
\begin{figure}
\includegraphics[width=3 in]{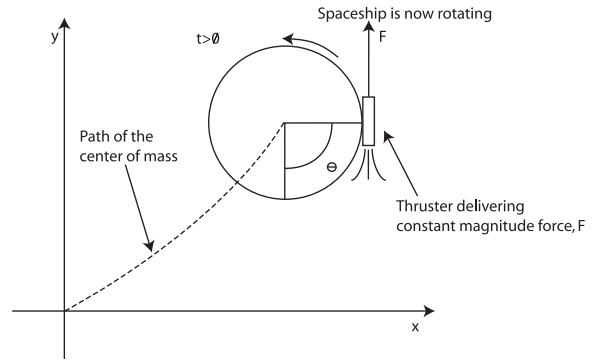}
\caption{\label{fig:initialmove} Qualitative sketch of the
position of the spaceship/thruster combination shortly after
$t=0$.  The ship is rotating and translating.}
\end{figure}

Three equations, one from the torque of
the thruster, and the other two from the $x$ and $y$ components of
the force of the thruster, define the motion. Let $\theta$ be the
angle the thruster has rotated about its center of mass since time
$t=0$. Initially, the ship will move in the $x$ direction and then
upward into the first quadrant as it begins to rotate as shown in
Fig.~\ref{fig:initialmove}.

From the torque on the system about the center of mass, the
rotational analog of Newton's second law ($\tau = I
\ddot{\theta}$) requires
\begin{equation}
FR = c\,mR^{2} \ddot{\theta}
\end{equation}
where the moment of inertia is $I=c\,mR^2$, where $c$ is a
dimensionless constant that depends on the distribution of mass
(e.g.~$\frac{1}{2}$ for a disk, but in general any positive
number), m is the mass of the ship, $R$ is the distance from the
center of mass to the thruster, and $\ddot{\theta}$ is the second
derivative of $\theta(t)$ with respect to time. Solving for the
angle as a function of time we have
\begin{equation}\label{thetat}
\theta(t)=\frac{Ft^2}{2c\,mR}
\end{equation}
assuming $\theta(0)=\dot{\theta}(0)=0$. The case of
$\dot{\theta}(0)\neq 0$ will be explored later in the paper.

Newton's second law requires
\begin{equation}
F \cos(\theta)=m\ddot{x}
\end{equation}\begin{equation}
 F \sin(\theta)=m\ddot{y}
\end{equation}
where $x$ and $y$ are the coordinates of the center of mass.
Substituting for $\theta$ and integrating these equations gives
velocity components:
\begin{equation}\label{vx}
v_{x}(t)=\frac{F}{m}\int_{0}^{t}\cos(\frac{Ft'^{2}}{2c\,mR})dt'
\end{equation}
\begin{equation}\label{vy}
v_{y}(t)=\frac{F}{m}\int_{0}^{t}\sin(\frac{Ft'^2}{2c\,mR})dt'
\end{equation}
assuming $v_{x}(0)=v_{y}(0)=0$. These integrals are well studied
and are called Fresnel integrals\cite{AandS}.  Their evaluation is
aided by a plot called the Cornu Spiral, which is shown in
Fig.~\ref{fig:cornu}.  Note the analysis required to this point
was within the level of the typical elementary calculus-based
physics course, though the integrals have led to special
functions.

\subsection{The motion of the center of mass and the Cornu Spiral}
\begin{figure}
\includegraphics[width=3 in]{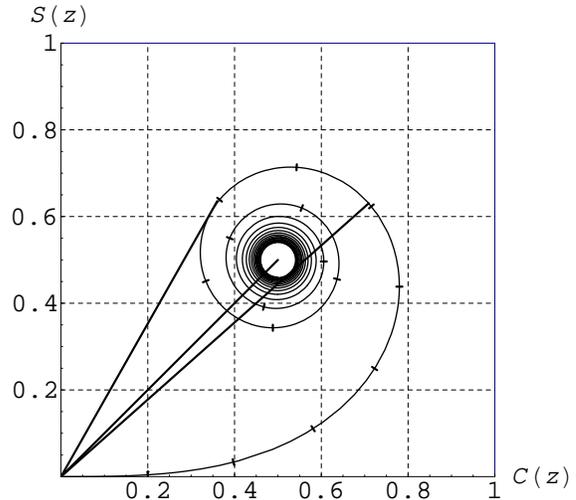}
\caption{\label{fig:cornu} The Cornu Spiral as it is often
shown\cite{hecht,AandS}. The Fresnel sine integral,
$S(z)=\int^{z}_{0} \sin(\frac{\pi}{2}t^2 ) dt$, is plotted on the
$y$-axis versus the Fresnel cosine integral, $C(z)=\int^{z}_{0}
\cos(\frac{\pi}{2}t^2 ) dt$, on the $x$-axis. The parameter $z$,
proportional to time for our problem, runs along the spiral with a
tick every $0.2$ stopping at 3.0. For the spaceship problem the
axes are proportional to the $x$ and $y$ velocity components. The
central portion of the spiral near $(0.5, 0.5)$ is not plotted for
clarity.  The three straight lines on the plot represent the
velocities with the steepest direction, the terminal speed, and
the maximum speed, respectively in a clockwise fashion.}
\end{figure}

Examining the Fresnel integrals for the velocity components
reveals much about the motion of the center of mass.  In the limit
$t\rightarrow \infty$, each component of velocity goes to
$\sqrt{\pi c F R/4 m}$.  Therefore, the center of mass moves off
at a $45$ degree angle with respect to the $x$-axis as time
approaches infinity. The Cornu Spiral, most commonly associated
with the problem of diffraction of a rectangular aperture
\cite{hecht}, represents these two integrals graphically.
Fig.~\ref{fig:cornu} is the Cornu Spiral as it is often displayed
\cite{AandS,hecht}. A point on this spiral (when multiplied by the
factor $\sqrt{\pi c F R/m}$) represents the $x$ and $y$ components
of velocity at a point in time. A line drawn from the origin to a
point on the spiral represents the instantaneous direction of the
motion of the center of mass.  Thus from this plot we see the
center of mass is always in the first quadrant since $v_{x}$ and
$v_{y}$ are always positive.  And since $v_{x}$ is positive at all
times then the plot of motion in the $y$ versus $x$ plane is
single valued.

Fig.~\ref{fig:cornu} also reveals a maximum speed (depicted by the
longest straight line) which is approximately equal to
$1.3422v_{\infty}$, where $v_{\infty}$ (straight line to center of
spiral) represents the speed of the center of mass as time
approaches infinity. The plot also shows a maximum angle with
respect to the $x$-axis for the trajectory approximately equal to
$60.466^{\circ}$, independent of any other parameters. Thus if the
thruster is fired for an appropriate finite period of time, one
could obtain a trajectory anywhere between zero and
$60.466^{\circ}$ with respect to the x-axis, or a terminal
velocity anywhere between zero and $1.3422v_{\infty}$.

\subsection{The path of the center of mass} Integrating the
velocity components, equations \ref{vx} and \ref{vy}, gives the
position of the center of mass:
\begin{equation}\label{pos x}
x(t)=\frac{F}{m}\int_{0}^{t}\int_{0}^{t''}\cos(\frac{Ft'^2}{2c\,mR})dt'
dt''
\end{equation}
\begin{equation}\label{pos y}
y(t)=\frac{F}{m}\int_{0}^{t}\int_{0}^{t''}\sin(\frac{Ft'^2}{2c\,mR})dt'
dt''
\end{equation}
where we have assumed $x(0)=y(0)=0$.  These integrals can be
evaluated numerically.  Interestingly, using integration by parts
the position can also be expressed analytically in terms of
equations \ref{vx} and \ref{vy}, the components of velocity. We
find
\begin{equation}\label{posxanal}
x(t)=-cR\sin(\frac{Ft^2}{2c\,mR})+t\,v_{x}(t)
\end{equation}
\begin{equation}\label{posyanal}
y(t)=cR[-1+\cos(\frac{Ft^2}{2c\,mR})]+t\,v_{y}(t).
\end{equation}
Thus the path of motion can be studied analytically through the
Fresnel special functions and its associated Cornu spiral.

\begin{figure}
\includegraphics{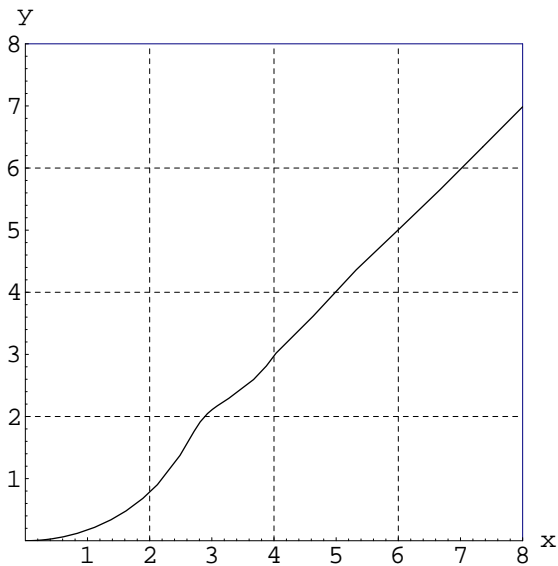}
\caption{\label{fig:motion} The path of the center of mass of the
spaceship. For actual distances the axes should be multiplied by
$c\,R$.  The shape of this path is universal and doesn't depend on
other parameters save for the scaling factor $c\,R$.}
\end{figure}

A plot of the motion is shown in Fig.~\ref{fig:motion}.  The shape
of the path is universal regardless of parameter values though
distances are scaled by the factor $c\,R$. Note that the
asymptotic trajectory as projected back towards the origin does
not pass through the origin but has a non-zero $x$-intercept.
Analytic analysis of the asymptotic ($t\rightarrow \infty$) forms
of equations \ref{posxanal} and \ref{posyanal} show this intercept
occurs $x=c\,R$. Curiously, while the thruster delivers equal $x$
and $y$ components of impulse (change in momentum) as time
approaches infinity to the center of mass, there is an asymmetry
in the displacement as shown by this intercept (due to $v_{x}$
being greater than $v_{y}$ initially).  For actual spacecraft
maneuvers we see a single thruster is not a very practical
configuration.

\section{Physical Intuition}
It has been our experience that the majority of students and
faculty alike have difficulty intuiting the motion of the center
of mass.  Physical intuition is not a well defined term.  An
interesting recent book entitled {\em Seeking Ultimates: An
Intuitive Guide to Physics} \cite{landsberg} states intuition is
something for a student ``to absorb in their bones.'' The
dictionary \cite{websters} defines intuition as
\begin{quote}
1a) the act or faculty of knowing without the use of rational
processes; immediate cognition b) knowledge acquired by use of
this faculty. 2.) acute insight
\end{quote}
We feel a definition of ``physical intuition'' requires more. A
recent article by Singh \cite{singh} agrees that physical
intuition is difficult to define but offers these words:
\begin{quote}
Cognitive theory suggests that those with good intuition can
effectively pattern-match or map a given problem onto situations
with which they have experience.
\end{quote}
These words provide a suitable footing for the term because below
we relate the problem at hand to a more common problem, which most
physicists have had experience with during the course of their
education.  Perhaps ``absorb in their bones'' is on the mark if
interpreted as absorbing a number of standard problems to provide
a bank with which to pattern match.

\subsection{Center of mass has a terminal velocity}
The simplest idea is that as the object spins faster and faster
the impulse to the center of mass over a single revolution must
tend to zero.  Therefore the change in linear momentum tends to
zero and thus the notion of a terminal velocity for the center of
mass is reasonable (though not guaranteed, the harmonic series
tends to zero but it's sum does not).

The first half of a revolution takes longer than the second and
thus it must be the case that the impulse is always positive in
the $y$ direction for any time and the motion is confined to the
upper half plane. In addition, $y$ plotted as a function of time
is monotonically increasing.  One may be tempted to draw similar
conclusions for the $x$ direction but here things are trickier,
especially for whether a plot of $x$ versus time is monotonically
increasing. To see this consider a slightly different problem.

\subsection{An alternate spaceship problem}
\begin{figure}
\includegraphics[width=3 in]{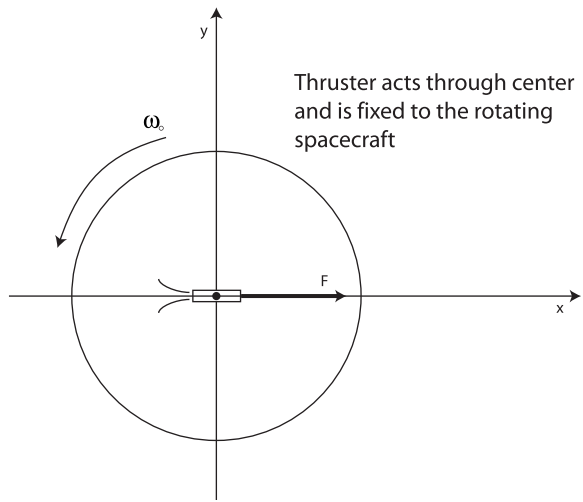}
\caption{\label{fig:alternate} Sketch of the initial position for
the alternate problem with the thruster acting through the center
of mass but with the ship initially rotating at angular velocity
$\omega_{o}$.}
\end{figure}

If the spaceship's thruster had acted through its center of mass
and had a rotation rate given by $\omega_{o}$, as pictured in
Fig.~\ref{fig:alternate}, then we could say something about the
$x$ component of velocity. During the first quarter of rotation
the $x$ component of acceleration is positive and the $x$
component of velocity goes from $0$ to some maximum. During the
second quarter of rotation the $x$ component of acceleration is
negative and the symmetry of the applied force dictates that this
acceleration will reduce the $x$ component of velocity to zero.
During the last half of the rotation the $x$ component of velocity
will be negative and the symmetry of the kinematics would return
the $x$ component of the center of mass to $x=0$. Then, as far as
the $x$ direction is concerned, the whole thing starts over again.
The overall motion, assuming $\theta(t) = \omega_{o} t$, is a
cycloid, reminiscent of the motion of a charged particle starting
at rest in orthogonal uniform electric and magnetic fields.

The velocity components are
\begin{equation}
v_{x}(t)=\frac{F}{m\omega_{o}}\mathrm{sin}(\omega_{o}t)
\end{equation}
\begin{equation}\label{vdrift}
v_{y}(t)=\frac{F}{m\omega_{o}}[1 - \mathrm{cos}(\omega_{o}t)]
\end{equation}
assuming $v_{x}=v_{y}=0$.  And the positions would be given by
\begin{equation}
x(t)=\frac{F}{m\omega_{o}^{2}}[1 - \mathrm{cos}(\omega_{o}t)]
\end{equation}
\begin{equation}
y(t)=\frac{F}{m\omega_{o}^{2}}[\omega_{o}t -
\mathrm{sin}(\omega_{o}t)].
\end{equation}

\begin{figure}
\includegraphics{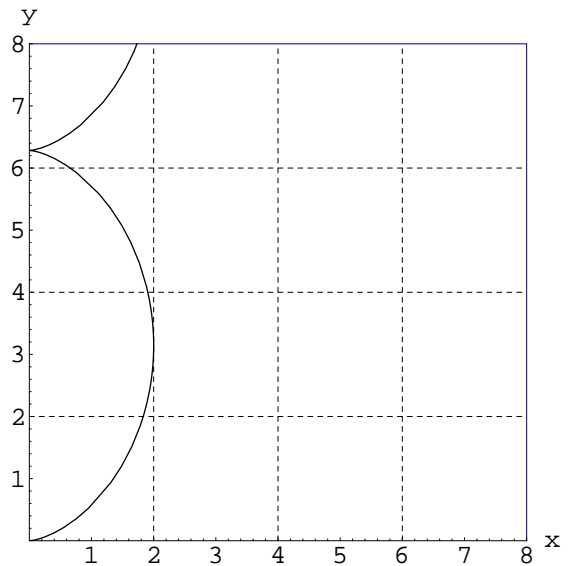}
\caption{\label{fig:cycloid} The cycloid motion of the center of
mass for the alternate problem.  The shape of the path is
universal.  The axes are in units of $F/m\omega_{o}^{2}$.}
\end{figure}

The path is depicted in Fig.~\ref{fig:cycloid} and the shape of
the path is also universal, though the axes are scaled by the
factor $F/m\omega_{o}^{2}$. One possible mistake is to confuse
constant rotation with uniform circular motion.  But uniform
circular motion is not a correct analogy since the force of the
thruster is not, in general, perpendicular to the velocity of the
center of mass.

\subsection{Spaceship is stuck in the first quadrant}
Returning to the original problem, the rotation rate is not
constant, but increases. As such we would expect the particle to
never return to $x=0$ since the time spent in each rotation
thrusting with a positive $x$ component will be longer than the
time spent with a negative $x$ component.  Thus the spaceship is
doomed to remain in the first quadrant for all its travels
contrary to a common misconception that the spaceship may move in
some sort of spiral around the origin.

\begin{figure}
\includegraphics{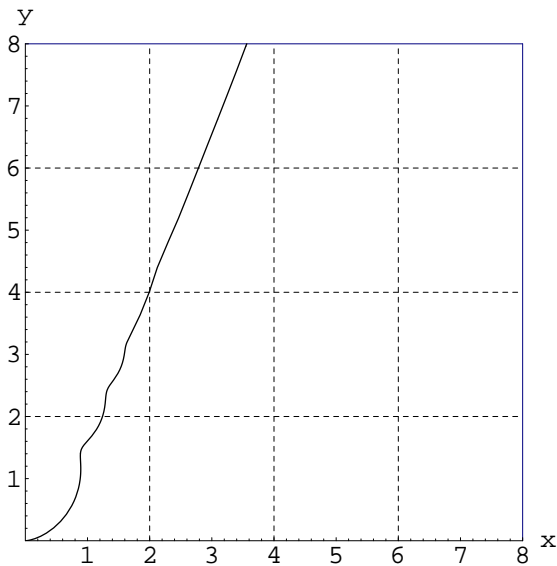}
\caption{\label{fig:initrot1} Path of motion of the center of mass
for a situation with initial rotation.  Again, as in
Fig.~\ref{fig:motion} the shape of this path is fixed for a given
value of $\omega_{o}^{2}cmR/F$, which in this case is $1$. The
axes are in units of $c\,R$.  Note the $x$ component of velocity
is negative for a period during the first rotation (near
$x=y=1$).}
\end{figure}

As mentioned earlier, we note the actual path of the center of
mass as described by a function $y(x)$ is single-valued, meaning
physically that the $x$ component of velocity (as well as the $y$
velocity component) is always positive. However, had there been an
initial rotation, $\omega_{o}^{2}cmR/F = 1$ for example, then
there would have been a negative $x$ component of velocity during
the first rotation and thus $y(x)$ would have been double valued
for some $x$ values as shown in Fig.~\ref{fig:initrot1}.  The
situation of a thruster with initial rotation is discussed in
detail below.

It is hoped the above discussion sheds some light on why the
$45$-degree asymptotic path, the non-zero $x$-intercept, and the
single-valued nature of $y(x)$ are difficult to intuit, even in
hindsight. They depend on the value of integrals that are not
intuitive (without the aid of the Cornu spiral or some other such
device).

We did succeed in providing an ``intuitive'' explanation to
explain that the path of motion is all in the first quadrant by
comparing to an alternative known elementary problem. And, we
intuited the notion of a terminal velocity and thus the asymptotic
path for large times is a straight line.

In the interest of full disclosure, we add that our initial
thoughts on the motion weren't always right, and we wrote this
section with the benefit of hindsight from solving the equations
of motion.

\section{Dimensional Analysis}
Dimensional analysis has been discussed, for example, in
association with models and data utilizing the simple pendulum as
an example \cite{price}, in a simple experiment involving the flow
of sand \cite{yersel}, and in the error analysis of a falling body
\cite{bohren}. This problem lends itself to dimensional analysis,
the most interesting example being the terminal velocity.  The
characteristic mass is $m$, length is $R$, and time is $\sqrt{m
R/F}$.  A fourth parameter for the problem is actually
dimensionless, the parameter $c$ from the form factor of the
moment of inertia.  Though dimensionless one can usually predict
whether a quantity should increase or decrease as a function of
$c$, though the power of the dependance on $c$ is unobtainable by
such analysis.

Table \ref{tab:table1} list a few quantities of possible interest,
such as terminal velocity ($v_{\infty}$) and the $x$ intercept of
the asymptotic path, along with the actual value and a dimensional
estimate. Note all numerical prefactors of the estimates are
within a factor of ten of the actual prefactor.

\begin{table}
\caption{\label{tab:table1} Quantities of interest, dimensional
estimates, and actual values for the problem with the thruster at
the edge of the spaceship.}
\begin{ruledtabular}
\begin{tabular}{lcr}
Quantity& Estimates & Actual\\
\hline
length & $R$ & --\\
mass & $m$ & --\\
time & $\sqrt{mR/F}$ & --\\
$t_{1st~rotation}$ & $\sqrt{mR/F}$ & $\sqrt{4 \pi c\,mR/F}$\\
$v_{\infty}$ & $\sqrt{FR/m}$ & $\sqrt{\pi cFR/2m}$\\
$x$ intercept of asymptotic path & $R$ & $cR$\\
displacement($t_{1st~rotation}$) & $R$ & $\sim 3.75cR$\\
\end{tabular}
\end{ruledtabular}
\end{table}

\subsection{Dimensional analysis of the alternate
 problem}
To physically understand the motion we introduced the alternate
problem of a spaceship initially rotating with thrust acting
through the center of mass.  This effectively eliminates the
radius, $R$, from the problem (since no torque is available the
rotation rate will not change), but it introduced a new parameter,
the initial rotation rate, $\omega_{o}$.  For this problem the
characteristic mass is $m$, length is $F/m \omega_{o}^{2}$, and
time is $1/\omega_{o}$.

Table \ref{tab:table2} is analogous to Table \ref{tab:table1} for
this alternate problem. Note again all numerical prefactors of the
estimates are within a factor of ten of the actual prefactor.

\begin{table}
\caption{\label{tab:table2} Quantities of interest, dimensional
estimates, and actual values for the alternate problem with the
thruster acting through the center of mass but with initial
rotation, $\omega_{o}$.}
\begin{ruledtabular}
\begin{tabular}{lcr}
Quantity & Estimates & Actual\\
\hline
length & $F/m\omega_{o}^{2}$ & --\\
mass & $m$ & --\\
time & $1/\omega_{o}$ & --\\
$t_{1st~rotation}$  & $1/\omega_{o}$ & $2 \pi/\omega_{o}$\\
$v_{drift}$ & $F/m\omega_{o}$ & $F/m\omega_{o}$\\
maximum $x$ & $F/m\omega_{o}^{2}$ & $2F/m\omega_{o}^{2}$\\
displacement($t_{1st~rotation}$) & $F/m\omega_{o}^{2}$ & $2\pi F/m\omega_{o}^{2}$\\
\end{tabular}
\end{ruledtabular}
\end{table}

\subsection{Original problem with initial rotation} If the original
problem had been initially rotating then there would have been two
length scales, two time scales, and even two mass scales.  The
second mass scale would be given by $F/R\omega_{o}^{2}$.  With two
sets of characteristic scales, dimensional analysis is of less
value because there are an infinite number of ways to construct
quantities of interest.  For example,  let $l_{1}, m_{1}, t_{1}$
be a characteristic length, mass and time respectively, and let
$l_{2}, m_{2}, t_{2}$ be a second set. Suppose we're curious about
a velocity.   Obvious possibilities are $l_{1}/t_{1}$ and
$l_{2}/t_{2}$.  But
\begin{equation}
\frac{l_{1}}{t_{2}},\hspace{0.1 in}
\sqrt{\frac{l_{1}l_{2}}{t_{1}t_{2}}},\hspace{0.1 in}
\frac{l_{1}m_{2}}{t_{1}m_{1}},\hspace{0.1 in}
\mathrm{or}\hspace{0.1 in} \sqrt{\frac{m_{2}}{m_{1}}} \left(
{\frac{l_{1}}{t_{1}}} \right)
\end{equation}
are examples of other possibilities.

Consider the initial problem but now allow an initial rotation
rate as well. The velocity components would be given by the
integrals:
\begin{equation}\label{vxwrot}
v_{x}(t)=\frac{F}{m}\int_{0}^{t}\cos({\frac{Ft'^{2}}{2c\,mR}}+\omega_{o}t')dt'
\end{equation}
\begin{equation}\label{vywrot}
v_{y}(t)=\frac{F}{m}\int_{0}^{t}\sin({\frac{Ft'^{2}}{2c\,mR}}+\omega_{o}t')dt'.
\end{equation}
Our intuition says the terminal velocity should decrease as
$\omega_{o}$ increases (for positive $\omega_{o}$, i.e.~in the
direction of the applied torque). Dimensional analysis for a
velocity reveals ambiguities such as:
\begin{equation}
\frac{F^{2}}{\omega_{o}^{3}cRm^{2}},\hspace{0.1 in}
\frac{F^{3/2}}{\omega_{o}^{2}c^{1/2}R^{1/2}m^{3/2}},\hspace{0.1in}
\frac{F}{\omega_{o} m}.
\end{equation}

\begin{figure}[h]
\includegraphics[width=3 in]{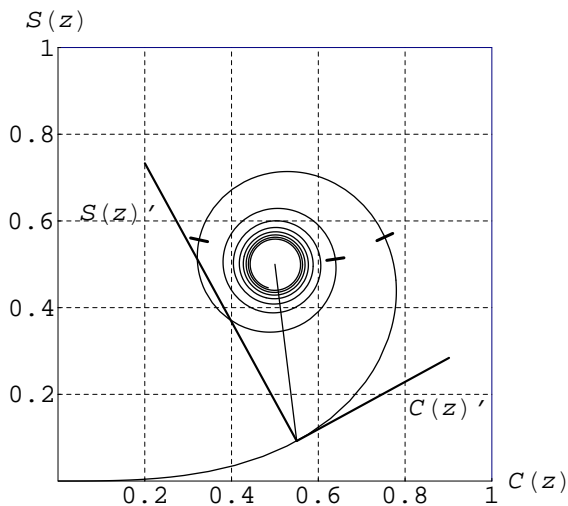}
\caption{\label{fig:cornushift} The Cornu spiral is shown with a
new axis placed along the spiral to account for an initial
rotation satisfying $\omega_{o}^{2}cmR/F=1$.  Note the trajectory
angle of the line which is proportional to $v_\infty$ (line from
the new origin to the point {$\frac{1}{2},\frac{1}{2}$}) is
parallel to the asymptotic trajectory in Fig.~\ref{fig:initrot1}
above. The ticks placed along the spiral represent other origins
corresponding to $\omega_{o}^{2}cmR/F = 4, 9$ and $16$.  As in
Fig.~\ref{fig:cornu} actual velocities are obtained by multiplying
the axes by the factor $\sqrt{\pi c F R/m}$.}
\end{figure}

The new velocity integrals can still be interpreted with the aid
of the Cornu spiral. By completing the square of the arguments of
the trigonometric functions, an initial rotation can be shown to
be a shift of $\sqrt{\omega^{2}_{o} c\,mR/\pi F}$ along the spiral
and a rotation of axis by an angle of $\omega^{2}_{o} c\,mR/2F$ as
shown in Fig.~\ref{fig:cornushift}, where the shifted axes are
placed for an initial rotation satisfying $\omega_{o}^{2}cmR/F=1$.

With the aid of this shifted axis, we see the terminal velocity
should indeed get smaller and also the angle of the trajectory
should increase.
 Also, both $v_{x}$ and $v_{y}$ should approach zero as $\omega_{o}$ approaches infinity. However,
since the trajectory tends to $90$ degrees from the $x$-axis as
$\omega_{o} \rightarrow \infty$ we note $v_{x}$ and $v_{y}$ cannot
tend to zero with the same dependence on $\omega_{o}$.

In fact, it can be shown that as $\omega_{o} \rightarrow \infty$
\begin{equation}v_{x}(t=\infty) \sim
\frac{F^{2}}{\omega_{o}^{3}cRm^{2}}-
\frac{3F^{4}}{\omega_{o}^{7}c^{3}R^{3}m^{4}} +{\cal
O}\left(\frac{1}{\omega^{11}_{o}}\right)
\end{equation}
and
\begin{equation}
v_{y}(t=\infty) \sim \frac{F}{\omega_{o} m}-
\frac{3F^{3}}{\omega_{o}^{5}c^{2}R^{2}m^{3}}+{\cal
O}\left(\frac{1}{\omega^{9}_{o}}\right).
\end{equation}
Each term in the expansions are further examples of the ambiguity
in constructing velocities with two characteristic scales.  The
asymptotic trajectory (angle from the $x$-axis) approaches $90$
degrees, since $\tan(\theta)=(v_{y}/v_{x}) \sim
\omega_{o}^{2}cRm/F$ as $\omega_{o} \rightarrow \infty$.

\section{Path of motion with a negative initial rotation}
\begin{figure}
\includegraphics{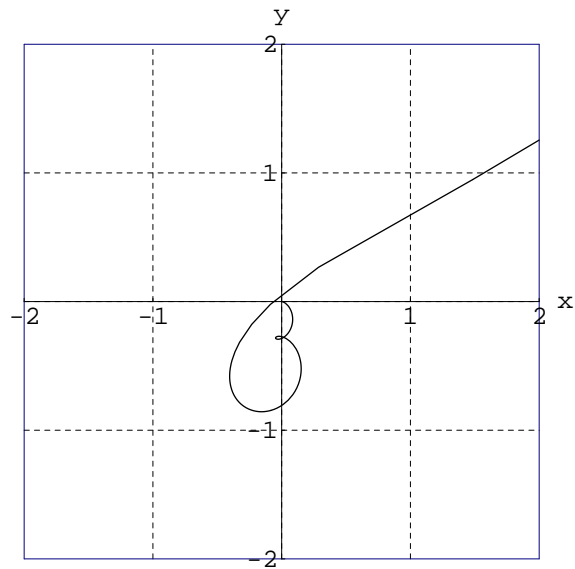}
\caption{\label{fig:initrotneg5} The path of the center of mass
for a situation with initial rotation opposite the applied torque
of the thruster, assuming $\omega_{o}^{2}cmR/F=25$. Axes are in
units of the factor $cR$.}
\end{figure}

\begin{figure}
\includegraphics{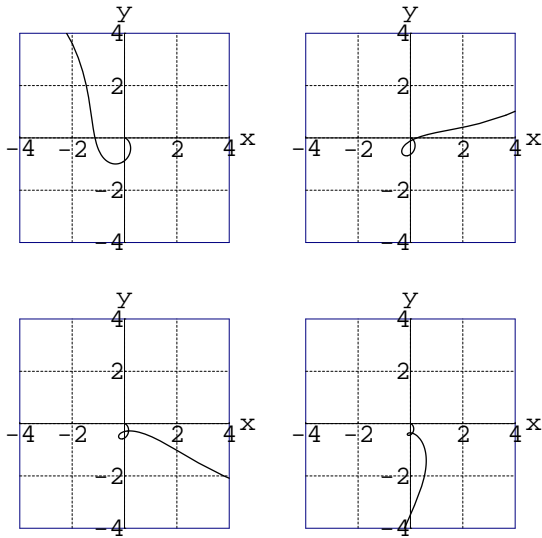}
\caption{\label{fig:initrotmulti} The path of the center of mass
for a situation with initial rotation opposite the applied torque
of the thruster, with $\omega_{o}=-3.2, -3.6, -3.8$ and $-4.2$
respectively in units of $\sqrt{F/cmR}$ (left to right then top to
bottom).  Axes are in units of the factor $cR$.}
\end{figure}

Since we have just generalized the original problem to include a
non-zero initial rotation aligned with the applied torque, it is
interesting to consider a negative initial rotation,
i.e.~initially spinning opposite the direction of the applied
torque. Fig.~\ref{fig:initrotneg5} displays the path of the center
of mass for the situation $\omega_{o}^{2}cmR/F= 25$ and with
$\omega_{o}$ being negative (i.e.~opposite the direction of the
torque). Notice that the displacement vector, for this case,
sweeps a polar angle somewhere between 270 and 360 degrees. This
raises questions: What is the maximum this angle could be? Could
the spaceship spiral
 around the origin, with an appropriate initial rotation, as some
incorrectly suggest for the original problem with no initial
rotation? Our explorations reveal that with an appropriate choice
of a negative $\omega_{o}$ the asymptotic path can be any compass
heading in the full 360 degree range of possibilities.
Fig.~\ref{fig:initrotmulti} shows four such possibilities
associated with four different initial rotation rates.

\subsection{Can the center of mass circle the origin? No}
As before, using integration by parts the components of position
can be expressed in terms of the velocity components.  We find:
\begin{equation}\label{posxanalwrot}
x(t)=-cR\sin(\frac{Ft^2}{2cmR}+\omega_{o}t)+(t+\frac{\omega_{o}cmR}{F})v_{x}(t)
\end{equation}
\begin{equation}\label{posyanalwrot}
y(t)=cR[\cos(\frac{Ft^2}{2cmR}+\omega_{o}t)-1]+(t+\frac{\omega_{o}cmR}{F})v_{y}(t)
\end{equation}
where $v_{x}(t)$ and $v_{y}(t)$ are those given by equations
\ref{vxwrot} and \ref{vywrot} respectively. For the spaceship to
circle the origin a necessary, but not sufficient, condition is
that $v_{y}=0$ while $y>0$.  Examining equation
\ref{posyanalwrot}, we see that if $v_{y}=0$ then
\begin{equation}
y\arrowvert_{v_{y}=0}=cR[\cos(\frac{Ft^2}{2cmR}+\omega_{o}t)-1]
\end{equation}
must be positive if we are to meet this condition.  But this
cannot ever be true since the most the cosine term could be is
$1$. Therefore once $y>0$ the spaceship will remain in the upper
half plane.  Note the above proof does not require any properties
of the Fresnel functions, just integration by parts.

\subsection{What is the displacement vector's maximum polar angle?}
\begin{figure}
\includegraphics{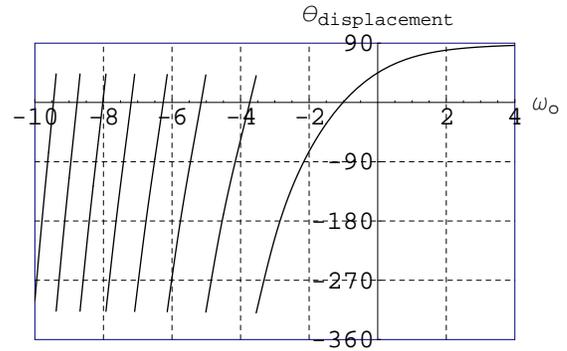}
\caption{\label{fig:maxangleplot} The polar angle swept by the
displacement vector (in degrees) from $t=0$ to $t=\infty$ as a
function of the initial rotation rate $\omega_{o}$ in units of
$\sqrt{F/cmR}$. The discontinuities correspond to paths that
intersect the origin. Thus, qualitatively, at the discontinuity
the path went from one like that in Fig.~\ref{fig:initrotneg5}
where it was going around the origin to one like that in the top
right of Fig.~\ref{fig:initrotmulti} where it cut back below the
origin. The maximum angles referred to in the text are defined as
negative angles on this plot.}
\end{figure}

Numerically, we have determined the maximum angle the displacement
vector can sweep while going around the origin is approximately
$319.52^{\circ}$ which occurs when the initial rotation rate is
approximately $\omega_{o}\approx -3.54$ in units of
$\sqrt{F/cmR}$. At approximately $\omega_{o}=-5.01$, which is near
that depicted in Fig.~\ref{fig:initrotneg5} the path again
approaches intersection with the origin with a corresponding
maximum polar angle for the displacement vector of approximately
$318.27^{\circ}$. There are infinitely many more of these pairs;
the list begins like this
\begin{eqnarray}
\matrix{ \omega_{o,1}\approx-3.54491& \hspace{0.3in} \theta_{1,
t=\infty}\approx319.522^{\circ} \cr \omega_{o,2}\approx-5.01326&
\hspace{0.3in} \theta_{2, t=\infty}\approx318.272^{\circ} \cr
\omega_{o,3}\approx-6.13996& \hspace{0.3in} \theta_{3,
t=\infty}\approx317.682^{\circ}  \cr \omega_{o,4}\approx-7.08982&
\hspace{0.3in} \theta_{4, t=\infty}\approx317.327^{\circ} \cr
\vdots \hspace{0.2in}& \hspace{0.6in} \vdots \hspace{0.3in}.}
\end{eqnarray}
The maximum polar angle appears to continue to decrease.  Thus
$319.52^{\circ}$ appears to be the approximate maximum regarding
encircling the origin.  A plot of the polar angle swept by the
displacement vector ($\theta_{displacement}$) from $t=0$ to
$t=\infty$ is shown versus the initial rotation ($\omega_{o}$) in
Fig.~\ref{fig:maxangleplot} in units of $\sqrt{F/cmR}$.

\section{Conclusion}
As one object (see appendix), one force problems go, this one may
rival the simple harmonic oscillator for its richness.  The
problem has utility in introducing a student to special functions
and handbooks such as Abromowitz and Stegun \cite{AandS}.  It also
provides a mechanical model for thinking about Fresnel integrals.
It could be used in a numerical methods course where comparison
between analytic (and analytic asymptotic) expressions versus
numerical techniques could be performed. The dimensional analysis
applied to this problem is useful for many other problems.  For
example, projectile motion possesses a universal path shape, a
parabola, characterized by the dimensionless parameter the launch
angle with the length scale set by $v_{o}^{2}/g$ where $v_{o}$ is
the initial velocity and $g$ the acceleration due to gravity.
When presenting a new problem to a student, a good question to ask
is to try to sort out how many sets of scales does the problem
encompass and what can dimensional arguments say about the answers
to any questions posed.  Finally, the problem is a challenging
test of physical intuition and it can be of interest to the
teacher and student alike to think about just what is meant by
such a term as ``physical intuition'' and how would one go about
improving it.

\section*{Acknowledgments}
The authors are indebted to Shane Burns, Brian Patterson, and
thoughtful referees for helpful comments and suggestions.

\appendix*
\section{The other object(s) - momentum conservation}
\begin{figure} [ht]
\includegraphics{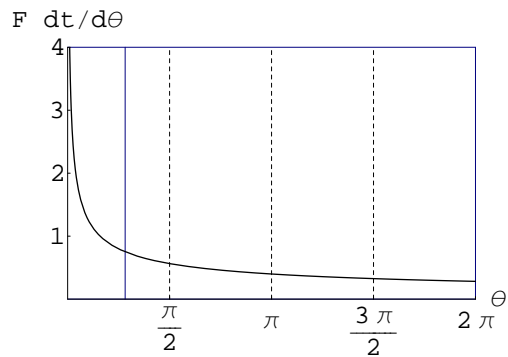}
\caption{\label{fig:impulse} The distribution of $F dt/d\theta$
during the first rotation as a function of the rotation angle. The
average over the first cycle occurs at $\sim 50.77^{\circ}$ ($\sim
0.886$ radians and denoted by a solid grid line), which is near
that of the asymptotic path of $45$ degrees.}
\end{figure}

The title stated there is only one body in this problem and indeed
the spaceship with its attached thruster has been our focus. But
momentum conservation suggests this cannot be the only thing in
our universe.  The thruster must be emitting something (perhaps a
photon) that carries momentum (and also energy and angular
momentum).  The momentum is carried away in all directions since
the spaceship rotates. The magnitude of the instantaneous impulse
imparted by the thruster, is $F dt$. The impulse per angular bin
from $\theta$ to $\theta + d\theta$ as a function of $\theta$
using Eq.~\ref{thetat} is then:
\begin{equation}
\frac{F dt} {d\theta} = \sqrt{\frac{F c\, m R}{2 \theta}}.
\end{equation}  A plot of this impulse density over the first cycle is
shown in Fig.~\ref{fig:impulse}.

Integrating the $x$-component, for example, over one revolution
(from $0$ to $2 \pi$) should be equivalent to evaluating
Eq.~\ref{vx} multiplied by $m$ from $t=0$ to $\sqrt{4 \pi c\, m R
/ F}$, i.e.
\begin{equation} F\int_{0}^{\sqrt{\frac{4 \pi c\, m R}{
F}}}\cos(\frac{Ft^{2}}{2c\,mR})dt = \int_{0}^{2 \pi}\sqrt{\frac{F
c m R}{2 \theta}}\cos(\theta) d\theta
\end{equation}

With the substitution $\theta = F t^{2}/2 c\, m r$ this is shown
to be true. In fact an alternative representation of the Fresnel
cosine integral is\cite{AandS}:
\begin{equation}
\int_{0}^{a}\cos(t^{2})dt =
\frac{1}{2}\int_{0}^{a^{2}}\frac{\cos(x)}{\sqrt{x}} dx.
\end{equation}

\vspace{0.2 in}

\vspace{2.5 in}

\bibliography{dudley}

\end{document}